\documentstyle[preprint,aps]{revtex}
\tighten
\preprint{HIP-1999-65/TH}
\begin{document}
\title{Quantum Magnetic Collapse}
\author{M. Chaichian $^{a,b}$, S.S. Masood$^{c,e}$, C. Montonen$^b$,\\
A. P\'erez Mart\'{\i}nez$^{d}$, H. P\'erez Rojas$^{a,b,e}$}
\address{\vskip 1cm 
$^a$ High Energy Division, Department of Physics, University of
Helsinki,\\
$^b$ Helsinki Institute of Physics,\\
P.O. Box 9, FIN-00014 University of Helsinki, Finland\\
$^c$ Physics Department, Quaid-i-Azam University, Islamabad, Pakistan\\
$^d$ ICIMAF, Calle E No. 309, 10400 La Habana, Cuba\\
$^e$ International Centre for Theoretical Physics\\
P.O. Box 586, Strada Costiera 11, 34100 Trieste, Italy}
\maketitle

\begin{abstract}
We study the thermodynamics of degenerate electron and charged vector boson
gases in very intense magnetic fields. In degenerate conditions of the
electron gas, the pressure transverse to the magnetic field $B$ may vanish,
leading to a transverse collapse. For $W$-bosons 
an instability arises because the magnetization diverges
at the critical field $B_c = M_W^2/e$. If the magnetic field is self-consistently maintained, 
the maximum value it can take is of the
order of $2 B_c/3$, but in any case the system becomes unstable and collapses. 
\end{abstract}

\vspace{1cm}


Large magnetic fields can be generated due to gravitational and rotational
effects in stellar objects like supernovas and neutron stars, i.e., magnetic
fields of order $10^{20}$G and larger have been suggested to exist in the
cores of neutron stars \cite{Chak}. The standard electroweak theory
establishes a limit on the magnetic field, the critical upper bound for
stable vacua being $B_c = M^2_{W}/e \simeq 1.06 \cdot 10^{24}$ G,
coming from the $W^{\pm}$ ground state
energy $\epsilon_{0q} = \sqrt{M_W^2 - eB }$, which is imaginary
for $B > B_c$. 
Fields of order $B_c$ may have been created at the
electroweak phase transition (see\cite{Vachaspati}, \cite{enqvist}). The
galactic and intergalactic magnetic fields can be considered as relics of
such huge magnetic fields in the early Universe \cite{29}-\cite{33}. In
astrophysics, also the critical field $B_{c^{\prime}}= m_e^2/e \simeq 4.41
\cdot 10^{13}$ G is relevant.

Nielsen, Olesen and Ambj{\o}rn \cite{Olesen},\cite{Ambjorn} showed
that the vacuum possesses the properties of a
ferromagnet or an antiscreening superconductor for $B \sim B_c$. It thus
seems relevant to study the electroweak medium in a strong magnetic field of
the order of the critical magnetic fields. The implications of these results
for astroparticle physics and cosmology are expected to be interesting.
As in preceding papers (Refs.\cite{yo},\cite
{polon}), we only consider the first generation of leptons and quarks
for the sake of simplicity. Here we shall calculate the magnetization due to
the charged leptons and intermediate vector bosons in the standard model.


The thermodynamic potential $\Omega = -T \ln {\cal Z}$ involves
contributions from leptons and quarks, which are
considered to be in chemical equilibrium among themselves through the boson
fields, described by equations among their chemical potentials \cite{yo}
like $\mu_{W^+} = \mu_{\nu} + \mu_{e^+}$, $\mu_{d_L} + \mu_{W^+} =
\mu_{u_L}$, $\mu_{e^+,W^+} + \mu_{e^-,W^-} = 0$. From the
thermodynamical potential we will choose the electron and $W$ sectors
exhibiting interesting effects in the astrophysical and cosmological
scenarios respectively in the presence of extremely strong magnetic fields ($%
B \sim B_{c^{\prime}}$ and $B \sim B_c$).


In the astrophysical scenario the electron-positron gas thermodynamics is of
interest.
In the cosmological context, we will be concerned especially with
the $W^{\pm}$ sector.

The one-loop thermodynamical potential per unit volume of the
electron-positron sector is $\Omega_e= \Omega_{se} + \Omega_{0e}$, where 
\begin{eqnarray}
\Omega_{se}& =& -\frac{eB}{4 \pi^2 \beta} \sum_{n = 0}^{\infty} a_n
\int_{-\infty}^{\infty} dp_3 \ln \left[ (1 + e^{-(E_q - \mu_e)\beta})(1 +
e^{-(E_q + \mu_e)\beta}) \right].
\end{eqnarray}
Here the sum extends over all Landau quantum numbers and the degeneracy
factor is $a_n = 2 - \delta_{0n}$, $E_q = \sqrt{p_3^2 + m_e^2 + 2eBn}$ and $%
\beta = T^{-1}$. For $W$'s, we have $\Omega_W= \Omega_{sW} + \Omega_{0W}$ 
\begin{eqnarray}
\Omega_{sW}& =& \frac{eB}{4 \pi^2 \beta} \int_{-\infty}^{\infty} dp_3 \ln
\left[ (1 - e^{-(\epsilon_{0q} - \mu_W)\beta})(1 - e^{-(\epsilon_{0q} +
\mu_W)\beta}) \right]  \nonumber \\[1em]
&&\mbox{} + \frac{eB}{4 \pi^2 \beta} \sum_{n = 0}^{\infty} b_n
\int_{-\infty}^{\infty} dp_3 \ln \left[ (1 - e^{-(\epsilon_q -
\mu_W)\beta})(1 - e^{-(\epsilon_q + \mu_W)\beta}) \right],
\end{eqnarray}
\noindent
where again we sum over all Landau quantum numbers and the degeneracy factor
is $b_n = 3 - \delta_{0n}$, with $\epsilon_{0q} = \sqrt{ p_3^2 + M_W^2 - eB }
$, and $\epsilon_q = \sqrt{ p_3^2 + M_W^2 + 2 eB (n + \frac{1}{2})}$.

The Euler-Heisenberg vacuum terms are, for the electron-positron field, 
\begin{equation}
\Omega_{0e} = \frac{e^2 B^2}{8\pi^2}\int_0^{\infty}e^{-m_e^2 x/eB}\left[%
\frac{\coth x}{x} -\frac{1}{x^2}-\frac{1}{3}\right]\frac{d x}{x},  \label{eul}
\end{equation}
\noindent
and for the charged gauge bosons, 
\begin{equation}
\Omega_{0W} = -\frac{e^2 B^2}{16\pi^2}\int_0^{\infty}e^{-M_W^2 x/eB}\left[%
\frac{1 + 2 \cosh 2x}{\sinh x} -\frac{3}{x}-\frac{7x}{2}\right]\frac{d x}{x^2},
\end{equation}
\noindent
which diverges at $B > B_c$, leading to a vacuum instability.

The mean density of particles minus antiparticles (average charge divided by 
$e$) is given by $N_{e,W} = - \partial \Omega_{e, W}/ \partial \mu_{e,W} $.
We assume that there is always a background charge of opposite sign, to
preserve electrical neutrality. We have 
\begin{eqnarray}
N_e & =&\frac{e B}{4 \pi^2 }\sum_0^{\infty}a_n \left[\int^{\infty}_{-\infty}
dp_3 (n_e^+ - n_e^-) \right],
\end{eqnarray}
\noindent
where $n_e^{\pm} = [\exp (E_{q} \mp \mu_e)\beta + 1]^{-1}$.

In the degenerate limit one gets $N_e = \frac{eB}{2 \pi^2}%
\sum_0^{n_{\mu}}a_n \sqrt{\mu^2_e - m^2 -2eBn}$, \noindent
where the integer $n_{\mu}= I[(\mu^2_e - m^2)/2eB]$.

For $W$-s, 
\begin{equation}
N_W =\frac{e B}{4 \pi^2 }\left[\int^{\infty}_{-\infty} dp_3 (n_{0p}^+ -
n_{0p}^-) \right] + \frac{e B}{4 \pi^2 }\sum_0^{\infty}b_n
\left[\int^{\infty}_{-\infty} dp_3 (n_p^+ - n_p^-) \right]
\end{equation}
\noindent
with $n_{0p}^{\pm} = [\exp (\epsilon_{0q} \mp \mu_W)\beta - 1]^{-1}$,
$n_{p}^{\pm} = [\exp (\epsilon_{q} \mp \mu_W)\beta - 1]^{-1}$.

The magnetization is given by the contribution of electrons and charged
vector bosons. It depends on the density of particles {\it plus}
antiparticles, and it is, 
\begin{equation}
{\cal M}_{W, e} = - \partial \Omega_{W, e}/\partial B
\end{equation}
where (calling ${\cal M}_{0e,0W}= - \partial \Omega_{0e,0W}/\partial B$), 
\begin{eqnarray}
{\cal M}_e & =&-\frac{\Omega_{se}}{B} - \frac{e }{4 \pi^2 }
\sum_0^{\infty}a_n \left[\int^{\infty}_{-\infty} dp_3 \frac{e B n}{E_q}
(n_e^+ + n_e^-)\right]+ {\cal M}_{0e},  \label{magne}
\end{eqnarray}
and in the degenerate limit \cite{polon}, 
\begin{equation}
{\cal M}_e = \frac{e }{4 \pi^2 }\sum_0^{n_\mu} a_{n}\left (\mu_e\sqrt{
\mu_e^2 - m^2-2e B n} - (m^2 + 4 e B n) \ln\frac{\mu_e + \sqrt{\mu_e^2 - m^2
-2 e B n}}{\sqrt{m^2 + 2 e B n}}\right) +{\cal M}_{0e},  \label{magne0}
\end{equation}
\noindent 
and 
\begin{eqnarray}
{\cal M}_W & =&-\frac{\Omega_W}{B} +\frac{e^2 B }{8 \pi^2 }%
\left[\int^{\infty}_{-\infty} \frac{dp_3}{\epsilon^0_q} (n_{0p}^+ +
n_{0p}^-) \right]  \nonumber \\[1em]
&&\mbox{} - \frac{e^2 B }{4 \pi^2 }\sum_0^{\infty} b_n (n + \frac{1}{2})
\left[\int^{\infty}_{-\infty} \frac{dp_3}{\epsilon_q} (n_p^+ + n_p^-)
\right] + {\cal M}_{0W}.  \label{magnw}
\end{eqnarray}


It is now especially interesting to discuss the equation of state
of the system. The total energy-momentum tensor, whose spatial diagonal
components are the pressures along the coordinate axes, may be obtained by
starting from the quantum statistical average $T_{\mu \nu
}=<{\cal T}_{\mu \nu }>_s$ where ${\cal T}_{\mu \nu }=\frac{\partial {\cal L}}
{\partial A_{\mu ,\nu }^a}A_{\mu ,\nu }^a-\delta _{\mu \nu }{\cal L}$ \cite
{Hugop}. If ${\cal L}$ is the total Lagrangian, after doing the statistical
average, its place in the energy-momentum tensor is taken by $\Omega $
(since $\Omega =-\beta ^{-1}\ln <e^{\int_0^\beta dx_4\int d^3x{\cal L}(x_4,
{\bf x)}}>_s$). In the present SU(2)$\times $U(1) model, the only non-zero averaged
components of the field tensor are those of the U(1)
external magnetic field tensor $F_{\mu \rho }$ and then, 
\begin{equation}
T_{\mu \nu }=(T\partial \Omega /\partial T+\mu_e \partial \Omega /\partial
\mu_e +\mu_W \partial \Omega /\partial \mu_W)\delta _{4\mu }\delta _{\nu 4}+
4F_{\mu \rho }F_{\nu \rho }\partial \Omega /\partial F^2-\delta _{\mu \nu
}\Omega.  \label{ten}
\end{equation}
For $F_{\mu \rho }=0$, (\ref{ten}) reproduces the usual zero field case \cite
{Hugop}. For the electrically charged particles, we obtain thus different
equations of state for directions parallel and perpendicular to the magnetic
field, 
\begin{equation}
p_3=-\Omega ,\hspace{1cm}p_{\perp }=-\Omega -B{\cal M}.  \label{tpr}
\end{equation}
This anisotropy in the pressures $p_3, p_{\perp }$ leads to a
magnetostriction effect in the quantum magnetized gas of charged particles.
If (\ref{ten}) is taken as the Maxwell stress tensor (classical case),
${\cal M} < 0$ and $p_{\perp }> p_3$, which produces a flattening effect in
white dwarfs and neutron stars models \cite{Shapiro}, \cite{Konno}. In the
present quantum case, for diamagnetic media also ${\cal M} < 0$ leading
again to a flattening effect. But for positive magnetization, the transverse
pressure exerted by the charged particles is smaller than the longitudinal
one by the amount $B{\cal M}$. The extreme case is found for magnetic
fields, $eB\gg T^2$, when the electrons are confined to the Landau ground
state $n=0$. (In what follows we will ignore the vacuum contribution to
electron-positron pressure and magnetization, which is justified at the
scale of densities and fields considered below). We have $\Omega_e =-B{\cal M}_e$
where, 
\begin{equation}
{\cal M}_e=\frac e{2\pi ^2}\left( \mu _e\sqrt{\mu _e^2-m^2}-m^2\ln \frac{\mu
_e+\sqrt{\mu _e^2-m^2}}m\right)  \label{magne1}
\end{equation}
\noindent 
and $\mu _e\simeq \sqrt{(2\pi ^2N_e/eB)^2+m^2}$, $N_e$ being the electron
density. As $\mu _e^2>m^2$, the expression (\ref{magne1}) is always positive
the system behaves as paramagnetic or ferromagnetic. But one of the most
important effects we have in this limit is that the transverse pressure
can vanish, 
\begin{equation}
p_{\perp }=-\Omega _e-B{\cal M}_e=0.  \label{Sam}
\end{equation}
\noindent
(This is the lower bound for the pressure. For fermions, the pressure cannot
be negative). The effect (\ref{Sam}) is of pure quantum origin and it is
easy to understand since all electrons are confined to the Landau ground
state, and the quantum average of their transverse momentum vanish. If we
consider a white dwarf star in which the predominating contribution to the
pressure is from the electron gas, the vanishing of $p_{\perp }$ means that
the gravitational pressure (of order $GM^2/R^4$ where $R$ is the geometric
average radius of the star) cannot be compensated and an instability appears
leading to a transverse collapse, i.e., the resulting object (a neutron star
or a black hole) would be ellipsoidal, in this case stretched along the
direction of the magnetic field, as a cigar.
It is interesting
to find the critical conditions for the occurrence of this confinement to
the state $n=0$, and in consequence, for the collapse. We have, 
\begin{equation}
n_\mu =I(\frac{\mu _e^2-m^2}{2eB})=\frac{2\pi ^4N_e^2}{e^3B^3}\sim
4.75\times 10^{-20}\frac{N_e^2}{B^3},  \label{minn}
\end{equation}
\noindent
and the condition $I(x)<1$ might be satisfied in some astrophysical
conditions. E.g., for $N_e\sim 10^{30}$, $B=3.36\times 10^{13}$ G, it is
enough that $B\gtrsim B_{c^{\prime }}$ to satisfy it. For densities of the
order of neutron stars, where a background of electrons and protons exist,
if $N_e=10^{39}$, the previous condition, if valid, would lead to
$B>10^{19}$G.


The $W$ population in the Landau ground state is significant if $d= \sqrt
{M^2_W - eB} \leq T$. In the degenerate limit, e.g. for $\sqrt{M^2_W+ eB}/T
\gg 1$, one can neglect the contribution from excited Landau states and by
taking only the $n=0$ term in (\ref{magnw}), one can approximate the first
two terms, since the main contribution to the integrals comes from very
small momenta,

\begin{equation}
{\cal M}_W = -\frac{eT}{4\pi}\sqrt{d^2-\mu^2_W}+\frac{eBT}{4\pi}\frac{1}
{\sqrt{d^2-\mu^2_W}}+ {\cal M}_{0W}.  \label{magnw2}
\end{equation}
The first term is the diamagnetic contribution which vanishes as $T \to 0$.
The third is the vacuum contribution, which is asymptotically 
\[
{\cal M}_{0W} \sim -\frac{2\Omega_0}{B}-\frac{e M_W^2}{16\pi^2}  \ln
(M_W^2/eB - 1), 
\]
\noindent
whose most important term is the second one which contributes para- or
ferromagnetically for $B>M^2/2e$, having a logarithmic divergence as $B \to
B_c$. As the logarithm is negative for $B_c/2 < B \leq B_c$, that term has a
negative contribution to the transverse pressure of vacuum for fields in
that interval. The first term of ${\cal M}_{0W}$ contributes diamagnetically.
But for $B \to B_c$ the dominant term in (\ref{magnw2}) is the second, which
is also para- or ferromagnetic, having a stronger divergence (inverse square
root) than the vacuum term. To have a more explicit form for (\ref{magnw2}),
one must write $\mu_W$ in terms of the charge density. When confined to the
Landau ground state the charge density of the system may be approximated as
($p_0$ is some characteristic momentum) 
\begin{equation}
N_W = \frac{e B T}{2 \pi^2 }\int_0^{p_0} \frac{d p_3}{\sqrt{p_3^2 + d^2 } -
\mu_W} -\int_0^{p_0} \frac{d p_3}{\sqrt{p_3^2 + d^2 } + \mu_W} \sim \frac{e
B T}{4 \pi }\frac{2\mu_W}{\sqrt{d^2-\mu^2_W}},  \label{5}
\end{equation}
\noindent
from which $\mu^2_W = d^2/[1 + \frac{e^2 B^2 T^2}{4\pi^2 N^2_W}]$, and 
\begin{eqnarray}
{\cal M}_W &=&-\frac{e^2 T^2 B d}{4\pi\sqrt{4\pi^2 N^2 + e^2 B^2 T^2}}+ 
\frac{e^2 B T}{4\pi d} \sqrt{1 + \frac{4\pi^2 N^2}{e^2 B^2 T^2}}  \nonumber
\\[1em]
&&\mbox{} + {\cal M}_{0W}.
\end{eqnarray}

Taking $N \geq 10^{39}$, $T \sim 10^{-8}$ ergs and $B \leq B_c$, one can
neglect the unity in the square root and contributions from the first
diamagnetic term in (\ref{magnw2}) and from ${\cal M}_0$, and one is left
with 
\begin{equation}
{\cal M}_W \simeq \frac{e N_W}{2 d}.  \label{d}
\end{equation}

The most important consequence is that the contribution of this
magnetization to the transverse pressure of the $W$ gas would be negative
(see (\ref{magnw})), and if ${\cal M}_W B$ contributes more than the
pressure of other species, (the partial pressure $p_3 = \Omega_W$ even
decreases as $B \to B_c$) an instability occurs since the total pressure
would be negative. Thus, for stability (also to prevent $W$ decay), we must
assume some background able to keep the total pressure $p_{\perp} \geq 0$.

Some sort of Bose-Einstein condensation
actually takes place \cite{conden} for bosons. For small momentum and
magnetic fields strong enough $B \sim B_c$, the term $1/d$ dominates and the
main contribution to the $W$ propagator comes from the low momentum gauge
bosons \cite{polon,conden}.

In the absence of a magnetic field, the quantum degeneracy of the $W$-boson
sector leads to condensation, which at $T \simeq 0$ has been estimated \cite
{Linde} to occur induced by neutrino densities of order $n_{c\nu
}=M_W^3/6\pi ^2\simeq 10^{45}{\rm cm}^{-3}$.
At any temperature, a spontaneous magnetization would appear in the condensate
of charged bosons, say $W^+$, even at zero external field $H = B -4\pi{\cal M%
} = 0$. This spontaneous magnetization could self-consistently maintain the
microscopic field $B =4\pi{\cal M}_{e,W}$.


Let us assume the magnetization large enough to maintain the internal field $%
B$ self-consistently and very large densities, such
that $\mu_e \gg m$. The dominant term in (\ref{magnw}) is (\ref{d}) $\sim e
N_W/2d$. At $B \sim B_c $ Gauss, we obtain that the self consistent critical
field is reached at an electron density $\sim 10^{48}$ electrons/cm${}^3$.

At such field intensities ${\cal M}_W$ diverges, but if we write the
self-consistency condition for the $W$ sector, we have 
\begin{equation}
B =4\pi{\cal M}= 2 \pi\frac{e N_W}{d}  \label{consis}
\end{equation}
Let us write $eB = x^2 M_W^2$ and since $0 \leq x \leq 1$, we easily get 
\begin{equation}
x^2\sqrt{1 - x^2} =\frac{2 \pi e^2 N_W}{ M_W^3} = A.  \label{cub}
\end{equation}
As $M_W^3/e^2 \sim 10^{49}$ cm${}^{-3}$, even for $N_W$ exceeding largely
the nuclear density, $A$ can be extremely small (For $A \sim 1$, $N_W \sim
10^{48} $ cm${}^{-3}$. The horizon of events is $\sim 5.6$ cm.) By writing $y
=x^2 \sqrt{1 - x^2}$ we have a curve having an increasing branch starting
from $x = 0$ up to a maximum at $x_M = \sqrt{2/3}$, $y = A_1 = 2\sqrt{3}/9$,
compatible with a density $N_W \sim 10^{48}$ cm${}^{-3}$. We have also a
decreasing branch from $x = x_M$ to $x=1$, compatible with densities smaller
than $10^{48}$ cm${}^{-3}$. Thus, eq.(\ref{cub} will not have real solutions
for $A > A_1$. However, for 
$A \leq A_1$ we have two real positive solutions for $x$ ( coinciding for $%
x^2 =2/3$). 
For $A \ll 1$, these solutions are $x_1 = \sqrt{A + A^2/2}$ and $x_2 = \sqrt{%
1 - A^2}$. The first solution means that $B$ increases with increasing $N_W$%
, (up to the value $B_M = 2 M_W/3e$). In the second solution $B$ decreases
as a function of $N_W$, its limit for $N_W \to 0$ being $B_c$. This
obviously indicates that the expression for the magnetization must include
the contribution from Landau states other than the ground state, which leads
to a diamagnetic response to the field. This would compensate the increase
of the self-consistent field with increasing $N_W$ to keep $B < B_{c}$.

This can be shown to occur from formula (\ref{magnw}). If we call the
ground state density $N_{Wg}$
and the density in other Landau states $N_{Wn}$ 
($N_W=N_{Wg}+\sum N_{Wn}$), for $B>B_M=2B_c/3$, $\partial B/\partial N_{Wg}<0$
and $\partial B/\partial N_{Wn}>0$ and excited Landau states start to be
populated. The condensate in the ground state decreases in favor of the
increase of the population in excited Landau states, which starts to grow
and contribute diamagnetically to the total magnetization keeping $4\pi{\cal M
}=B<B_c$. But for the system to react in this way, an enormous amount of
energy (and angular momentum) would be required, of the order of respectively
$N_WM_W$ and $N_W$ (here we neglect the running of $M_W$). But the transverse
collapse takes place at such densities: since  the pressure comes essentially from the
fermion (electron) background, the self-consistency condition
$B=4\pi {\cal M}_{e,W}$ leads to $p_3=-\Omega \sim B{\cal M}_e$ and $p_{\perp
}= p_3 -B({\cal M}_e + {\cal M}_W) \lesssim 0 $ and thus the system collapses.

Let us assume that in some stage of the early universe a very large external
field $H \sim B_c$ was present.
If $T \sim M_W$,
as happened near and below the electroweak phase transition, using up the
energy and angular momentum of the background radiation, $W^{\pm}$ pairs
will be produced in the energetic more favourable Landau ground state (having
a "mass" $= d$), and this process would be even more favoured as the magnetic
field approaches $B_c$ even for lower temperatures. The magnetization
${\cal M}$ is given by an expression similar to the second term in (\ref{magnw}),
in which the expressions for the particle-antiparticle densities would be in
equilibrium with the electromagnetic background radiation. This means that
one must take the chemical potential as zero (equal number of $W^{\pm}$).
Then, ${\cal M}_W \sim e^2 B T/4 \pi^2 d$. The density of particle +
antiparticle pairs would then be $\simeq eBT \sim 10^{48}$ cm$^{-3}$, and
the microscopic field $B < B_c$ starts to be maintained self-consistently. We
would have the situation discussed in the previous paragraph. The process of 
$W$ pair creation in the external field would lead to a creation of order
from disorder, i.e., to an effective cooling of the subsystem considered,
although due to similar reasons as before, $p_{\perp}\lesssim 0 $ and the
system becomes unstable and collapses.


We conclude, first, that if a degenerate electron gas is confined to its
Landau ground state, its transverse pressure vanishes. This phenomenon
establishes a limit to the magnetic fields expected to be observable in
white dwarfs, and even in neutron stars. Second, that the instability of the
vacuum in magnetic fields $B \sim B_c$ in a hot and dense medium, is
avoided, since the self-consistent magnetization prevents fields greater than
$2B_c/3$, although under such conditions the system becomes also unstable and collapses.

\section*{Acknowledgments}

Three of us (S. S. M., A. P. M. and H. P. R.) would like to thank Professor
Virasoro, the ICTP High Energy Group, IAEA and UNESCO for hospitality at the
International Centre for Theoretical Physics. H.P.R. thanks A.E. Shabad
for a discussion and the University of Helsinki for hospitality. The partial
support by the Academy of Finland under Project No. 163394 is greatly
acknowledged.

\end{document}